\documentclass[aps,twocolumn,showpacs,amsmath,amssymb]{revtex4-1}
\usepackage{float}
\usepackage{amsmath}
\usepackage[dvips,final]{graphicx}
\usepackage{epsfig}
\usepackage{color}
\begin{document}

\title{Magnetic anisotropy of Fe$_{1-y}$X$_y$Pt--L1$_0$~[X=Cr,Mn,Co,Ni,Cu] bulk alloys.}

\author{R. Cuadrado,$^1$ Timothy J. Klemmer,$^2$ and R. W. Chantrell$^1$}

\address{$^1$Department of Physics, University of York, York YO10 5DD, United Kingdom\\ 
$^2$Seagate Technology, Fremont, California, 94538, USA}

\date{\today}

\begin{abstract}
We demonstrate by means of fully relativistic first principles calculations 
that, by substitution of Fe by Cr, Mn, Co, Ni or Cu  in FePt--L1$_0$ bulk 
alloys, with fixed Pt content, it is possible to tune the magnetocrystalline 
anisotropy energy by adjusting the content of the non--magnetic species in 
the material. The changes in the geometry due to the inclusion of each element 
induces different values of the tetragonality and hence changes in the magnetic 
anisotropy and in the net magnetic moment. The site resolved magnetic moments 
of Fe increase with the X content whilst those of Pt and X are simultaneously 
reduced. The calculations are in good quantitative agreement with experimental 
data and demonstrate that models with fixed band structure but varying numbers 
of electrons per unit cell are insufficient to describe the experimental data 
for doped FePt--L1$_0$ alloys.   
\end{abstract}

\pacs{}

\maketitle

The chemically ordered face--centred tetragonal~({\it fct}) FePt--L1$_0$ 
phase has attracted much interest because of its large magnetocrystalline 
anisotropy energy~(MAE) value of 7$\times$10$^7$~ergs/cm$^3$ and hence 
its potential applications such as ultra--high density magnetic recording 
media~\cite{weller,ivanov}. From the experimental side there are two principal 
challenges to the production of high density recording media. First is the large 
values of the MAE to overcome the superparamagnetic limit so as to avoid the loss 
of recorded information~\cite{Yan} arising from thermal instability. One solution 
to this problem is the use of L1$_0$ bimetallic alloys as magnetic recording 
media. This leads to a second challenge in that the preparation of such alloys 
generally leads to the deposition of the disordered {\it fcc} A1 phase with low 
anisotropy. Transformation into the L1$_0$ phase with large MAE requires elevated 
annealing temperature, leading to problems with maintaining the granular structure 
necessary for high density recording.

Some studies have proposed using FePt--based ternary alloys to lower the kinetic 
ordering temperature leading to reduced annealing temperatures thereby improving 
the orientation and granular structure~\cite{yang,bwang,bwang1}. However, this has 
the detrimental effect of reducing the MAE. This problem was studied in a related 
theoretical paper by Sakuma~\cite{sakuma}, who used a fixed band structure corresponding 
to FePt and varied the number of electrons/unit cell $n_{eff}$, finding that the MAE 
had its maximum value for $n_{eff}=8$, corresponding to pure FePt. These predictions 
were verified experimentally by Suzuki et.al.~\cite{suzuki}. More recent work has 
studied the effect of substituting Fe by Cu~\cite{gilbert}, Mn~\cite{Xu,meyer,burkert}, 
Ni~\cite{Berry,Thiele}. Gilbert et. al.\cite{Thiele} conclude that Cu doping gives a 
relatively simple approach to achieve high quality L1$_0$ FeCuPt films that have greater 
MAE values than current media and therefore are desirable for future magnetic recording 
technologies. Also, the magnetic properties can be smoothly tuned by Cu-substitution 
into Fe sites of the ordered alloy. The experiments are generally supported by the 
rigid band model calculations of Sakuma~\cite{sakuma}. However, it is debatable as to 
how realistic such models are, given the chemical and structural changes induced by
alloying. This question can only be answered by a detailed investigation taking account 
of the nature of the alloys produced. This is the aim of the current letter. Our 
calculations are in good agreement with experiment and demonstrate a doping--species 
dependence of the MAE reduction and also, importantly, variations in the local MAE arising from the different lattice sites available to the impurity atoms.
    
Consistent with the experimental studies we proceed by replacing the Fe content in bulk 
FePt--L1$_0$ by Cr, Mn, Co, Ni and Cu keeping the Pt concentration fixed. Within 
Fe$_{1-y}$X$_y$Pt bulk alloy, we take the $y$ concentration  as 0, 0.25, 0.50, 
0.75 and 1, $y$ being the amount of non--magnetic species, whose introduction changes  
the effective number of valence electrons, $n_{eff}$, in the cell. The effective 
valence electrons are computed as $\sum_sN_s\cdot Z_{val}^s/N_{tot}$, where $N_s$ is 
the number of atoms of each species, $s$, and $N_{tot}$ the total number of atoms 
in the simulation cell. This systematic Fe replacement of similar 3$d$ elements 
serves to control $n_{eff}$ since for the above mentioned species, the valence charge 
$Z_{val}^s$ is 6, 7, 9, 10, 11, respectively. Note that only the 3$d$$s$ electrons are 
included in the current $n_{eff}$ definition; the 5$d$ contribution from the Pt is 
constant and is not taken into account here. 

All the geometric, electronic and magnetic structure calculations of 
Fe$_{1-y}$X$_y$Pt--L1$_0$ alloys have been done by means of DFT
using the SIESTA~\cite{siesta} code. As exchange correlation~(XC) 
potential we have employed the generalized gradient approximation~(GGA) 
following the Perdew, Burke, and Ernzerhof~(PBE) version~\cite{pbe}. 
To describe the core electrons we have used fully separable 
Kleinmann-Bylander~\cite{kb} and norm-conserving pseudopotentials
(PP) of the Troulliers-Martins~\cite{tm} type. As a basis set, we 
have employed double-zeta polarized~(DZP) strictly localized numerical 
atomic orbitals~(AO). The so--called electronic temperature --kT in the 
Fermi-Dirac distribution-- was set to 50~meV. The magnetic anisotropy 
energie~(MAE) has been obtained using a recent fully relativistic~(FR) 
implementation~\cite{LS-paper} in the GREEN~\cite{green,greenp} code 
employing the SIESTA framework. As usual, the MAE is defined as the 
difference in the total energy between hard and easy magnetization 
directions. Convergence of the MAE convergence  is dependent on the 
sampling $k$ points. Within the present work we performed an exhaustive 
analysis of the MAE convergence in order to achieve a tolerance below 
microelectron volts. We employed more than 5000 $k$ points in the 
calculations for each geometric configuration, which was sufficient to 
achieve the stated accuracy. 

\begin{figure}[b]
 \includegraphics[scale=0.40]{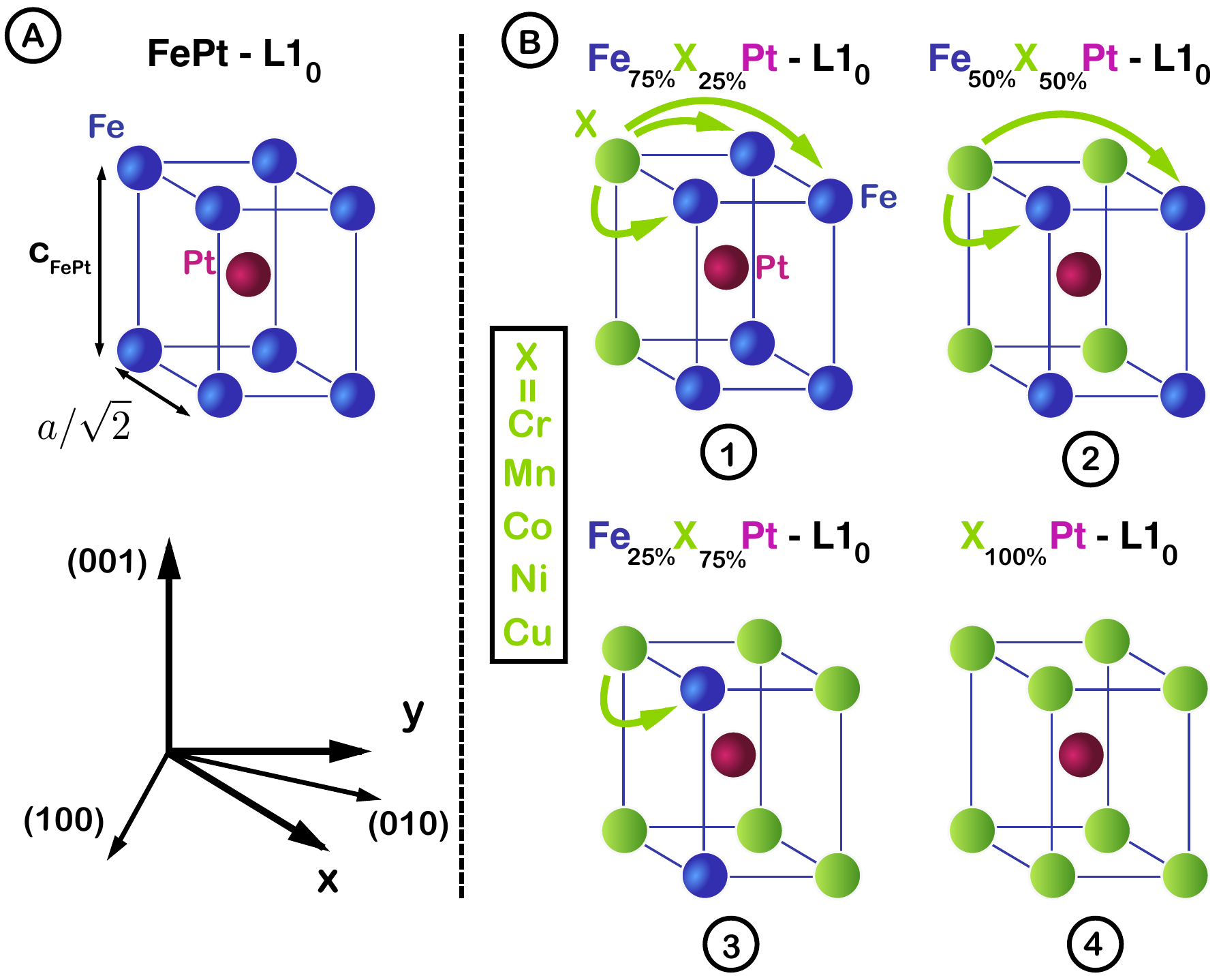}
 \caption{(Color online) (A) Schematic picture of the FePt--L1$_0$ unit cell 
          and its characteristic lattice values: $a$ and $c/a$. Notice that 
          the in--plane diagonal of the unit cell corresponds to the lattice 
          constant whilst the edge is $a/\sqrt{2}$; (B) Fe$_{1-y}$X$_y$Pt unit 
          cells. For $y$ values of 0.25, 0.50, 0.75 or 1, the cells have been 
          labeled from 1 to 4, respectively. The green arrows represent 
          different locations for the X species~(see text for explanation). 
          For specific alloys the X element will be one of Mn, Cr, Co, 
          Ni and Cu.}
 \label{geom-fig}
\end{figure}

The binary L1$_0$ alloys are formed by alternating planes of two distinct species 
which generates a vertical distortion as a result of which two quantities define 
the geometric structure: the in--plane lattice constant, $a$, and the out--of--plane 
parameter, $c$. Specifically, in FePt--L1$_0$ the experimental values are $a_{FePt}$=
3.86\AA\ and $c/a$=0.98. What we pursue is to study the variation of the anisotropy 
of bulk FePt--L1$_0$ via the substitution of Fe atoms by other 3$d$ species keeping 
Pt fixed. In doing this we are able to scan two possible ways to control the MAE: 
on the one hand, the species and on the other, the concentration of the impurity~(See 
Fig.~\ref{geom-fig}). Each one of the Cr, Mn, Co, Ni and Cu atoms has different number 
of valence electrons, so that it gives the possibility to control the number of total 
valence electrons in the cell depending on whether one, two, three or four atoms are 
replaced on the Fe sites. 

It is complicated in DFT to deal with this kind of calculation due to the large cells 
that one has to use to have a good approximation of the real material in a computer 
model, so we doubled the unit cell in X, Y and Z axis in order to be able to substitute 
the X atoms one by one. The minimum unit cell for a XPt--L1$_0$ bulk is composed of 
two atoms~(see Fig.~\ref{geom-fig}--B) and in our case the simulation cell has 16. 
This permits to move X atoms on different in--plane and out--of--plane positions as 
the green arrows depict in Fig.~\ref{geom-fig}--B. The number of different geometric 
configurations keeping the X content fixed were: three for B--1 and B--3 and six for 
B--2. For each configuration we performed a fully relaxation using the conjugate 
gradient~(CG) method without any constraint. Special attention is needed for the 
Fe(Cr,Mn)Pt alloys inasmuch as the lower energy configuration corresponds to 
antiferromagnetic~(AF) alignment of the Mn atoms between different atomic planes~\cite{lu}, 
so we include this restriction in our calculations as a magnetic constraint. 

\begin{figure}
 \includegraphics[scale=0.23]{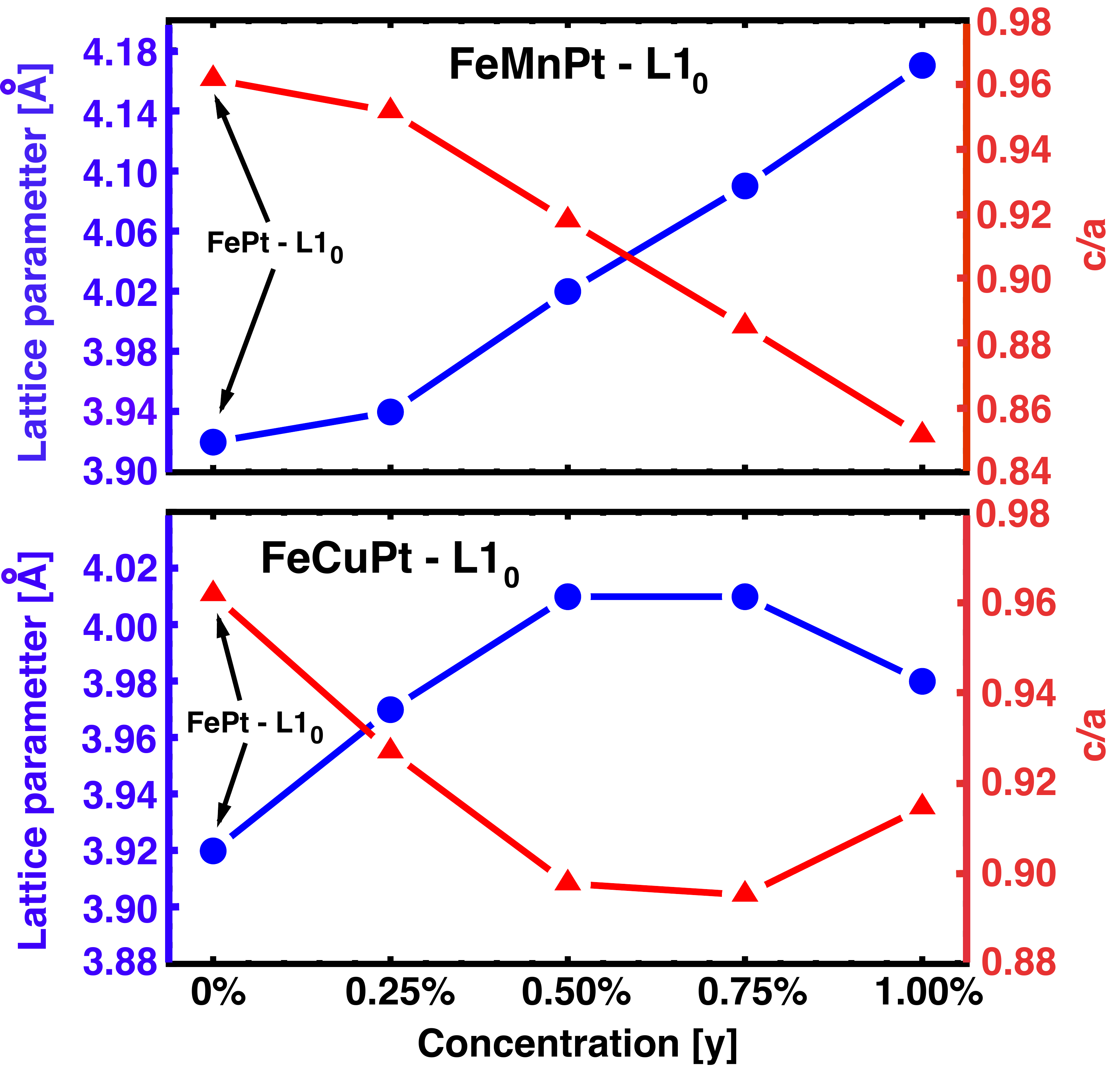}
 \caption{(Color online) Lattice constant parameters, $a$ and $c/a$, blue circles
          and red triangles, respectively, as a function of the $y$ concentration 
          for Fe$_{1-y}$(Mn,Cu)$_y$Pt--L1$_0$ bulk phases. The straight lines 
          are guide for the eye.}
 \label{latt-fig}
\end{figure}

Geometric, electronic and magnetic properties have been calculated by means of 
the mean value of fixed X:Fe composition unless explicitly expecified, for 
example, the density of states~(DOS) for a fixed X:Fe ratio $\alpha$  is
\[
DOS^{\sigma,\alpha}(\epsilon,X)=\frac{\sum^{N^\alpha}_{j=1}DOS^{\sigma,\alpha}_j(\epsilon,X)}{N^\alpha}
\]
where $j$ runs up to the total number of configurations for a fixed X composition, 
$N^\alpha$, $\sigma$ is the spin--up/--down states and $\epsilon$ the energy, usually 
shifted to the Fermi level, $\epsilon=E-E_F$.  

As was pointed out in the experimental work of Gilbert {\it et al}~\cite{gilbert}, 
the substitution of Cu into the bulk FePt--L1$_0$ phases promotes an increase of 
the in--plane lattice parameter $a$ values as we observe in Fig~\ref{latt-fig}~(blue 
dots), which is therefore in agreement with experiment. The out--of--plane $c$ 
parameter is simultaneously reduced with increasing Cu content. Consequently, 
the tetragonal distortion of FeCuPt--L1$_0$ increases with Cu content leading to 
an in--plane~(out--of--plane) lattice constants of 3.98\AA(3.64\AA). In the case 
of FeMnPt--L1$_0$(AF), $a$ increases with decreasing $c$, in agreement with the 
experimental results of Meyer~{\it et al}\cite{meyer}.   
 
\begin{figure}[b]
 \includegraphics[scale=0.15]{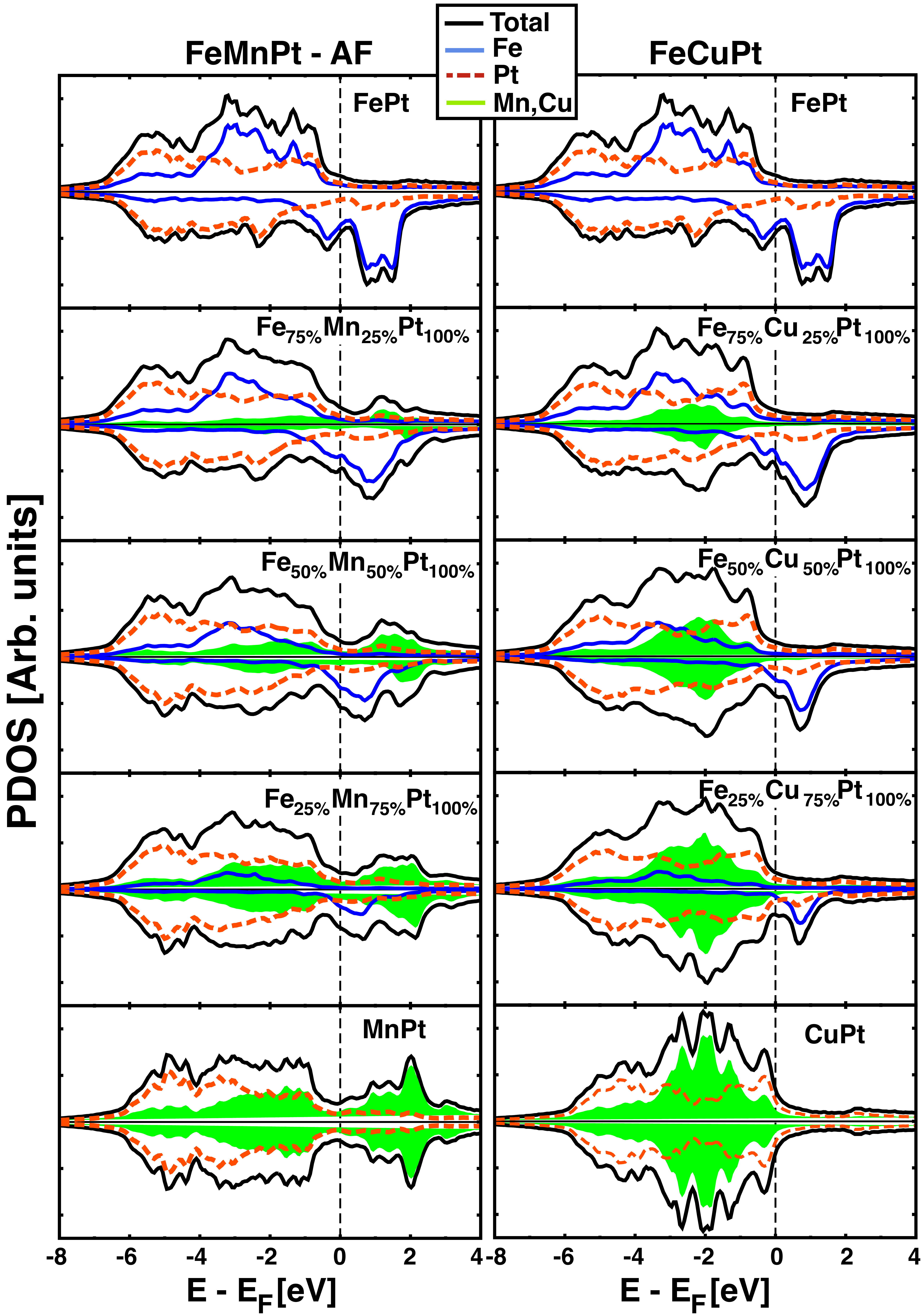}
 \caption{(Color online) Spin resolved density of states of 
          Fe$_{1-y}$(Mn,Cu)$_y$Pt~[y=0,0.25,0.50,0.75 and 
          1], left and right, respectively. The total DOS~(black 
          solid line) has been splitted in its Fe~(blue solid 
          line), Pt~(red dashed line) and Mn, Cu~(green filled 
          curve) contributions. The first two graphs on top,
          represent the DOS for the pure FePt--L1$_0$ bulk 
          alloy.}
 \label{PDOS-fig}
\end{figure}

Fig.~\ref{PDOS-fig} shows the averaged spin resolved density of states 
for the Fe$_{1-y}$(Mn,Cu)$_y$Pt alloys as the concentration $y$ of the 
Fe and (Mn,Cu) changes, left and right panel, respectively. From top to 
bottom is shown the evolution of the total~(solid black),  Fe~(solid blue), 
Pt~(dashed red) and the X(=Mn,Cu)~(filled green) DOS as the Fe, Mn, Cu species 
is varied. As  pointed out earlier, the lower energy configuration for the 
FeMnPt--L1$_0$ corresponds to AF coupling of the Mn atoms on alternating planes 
so that the up/down charges are equal and the net magnetic moment~(MM) is zero 
which is reflected in the green DOS curves. 

\begin{figure}[t]
 \includegraphics[scale=0.25]{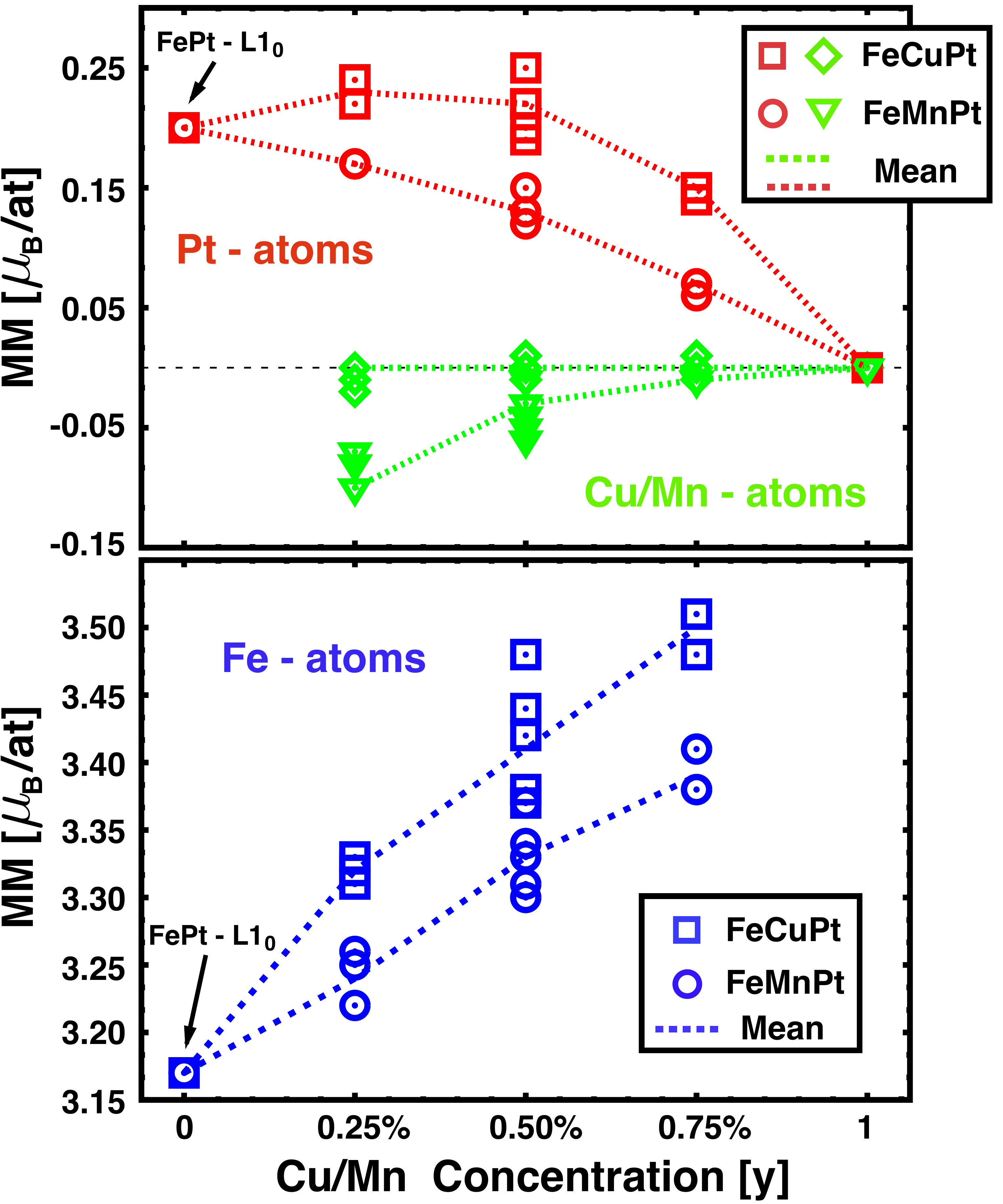}
 \caption{(Color online) Site resolved magnetic moment for Fe$_{1-y}$(Mn,Cu)$_y$Pt 
          alloys as a function of the Cu/Mn concentration $y$. In each panel has 
          shown separately the MM of the non--magnetic species~(upper) and the MM 
          of the Fe~(bottom). For the same alloy, fixing the Cu/Mn concentration, 
          similar symbols represent all the studied configurations~(see Fig.~\ref{geom-fig}). 
          The dashed lines refer to the mean value, $\sum_j$MM$_{j,X}^i$/N$_{conf}^i$,
          where X is Fe, Mn, Cu or Pt and $i$ refers some particular concentration.}
 \label{MMs-fig}
\end{figure}

The DOS curves aid the interpretation of the behavior of the magnetic moments. 
In bulk FePt--L1$_0$, the net MM is 3.37~$\mu_B/f.u.$, mainly dominated by the 
Fe species as depicted the blue line in the upper graphs. Only a fraction of 
this net value is contributed by the Pt sites, as has been pointed out in previous 
work~\cite{mryasov}. The Fe--Pt--Fe indirect interactions promote the polarization 
of the Pt atoms. In our case, the substitution of the the Fe by  non--magnetic 
species such as Mn or Cu reduces principally the Fe down--states tending to 
leave the Fe atoms embeded in a non--magnetic environment, becoming almost 
magnetically isolated with increasing Mn/Cu content. This is the reason behind 
the increase of MM$_{Fe}$  with Cu/Mn content~(see Fig.~\ref{MMs-fig}).
Simultaneously, the MM$_{Pt}$ diminishes due to the reduction in Fe neighbors 
until the up and down charges compensate. The MM$_{Cu}$ is close to zero 
independent of the amount present in the alloy. On the contrary, a small 
concentration of Mn gives a MM$_{Mn}$ value of -0.12$\mu_B/at$, its magnitude 
reducing to zero with increasing Mn content. In summary, both the (Mn,Cu)Pt--L1$_0$ 
bulk alloys have a zero net MM as we see in the bottom panel. The addition of the 
Fe atoms to these alloys enhances the value of the total MM, partly from the Fe 
and partly from the induced Pt polarization. It should be noted that in 
Fig.~\ref{MMs-fig} the scatter of points for a given impurity concentration indicates 
the variation of the magnetic moment across the different lattice positions hosting 
the impurity. A similar dispersion is also seen in the local MAE values, which are 
considered next.

\begin{figure}
 \includegraphics[scale=0.25]{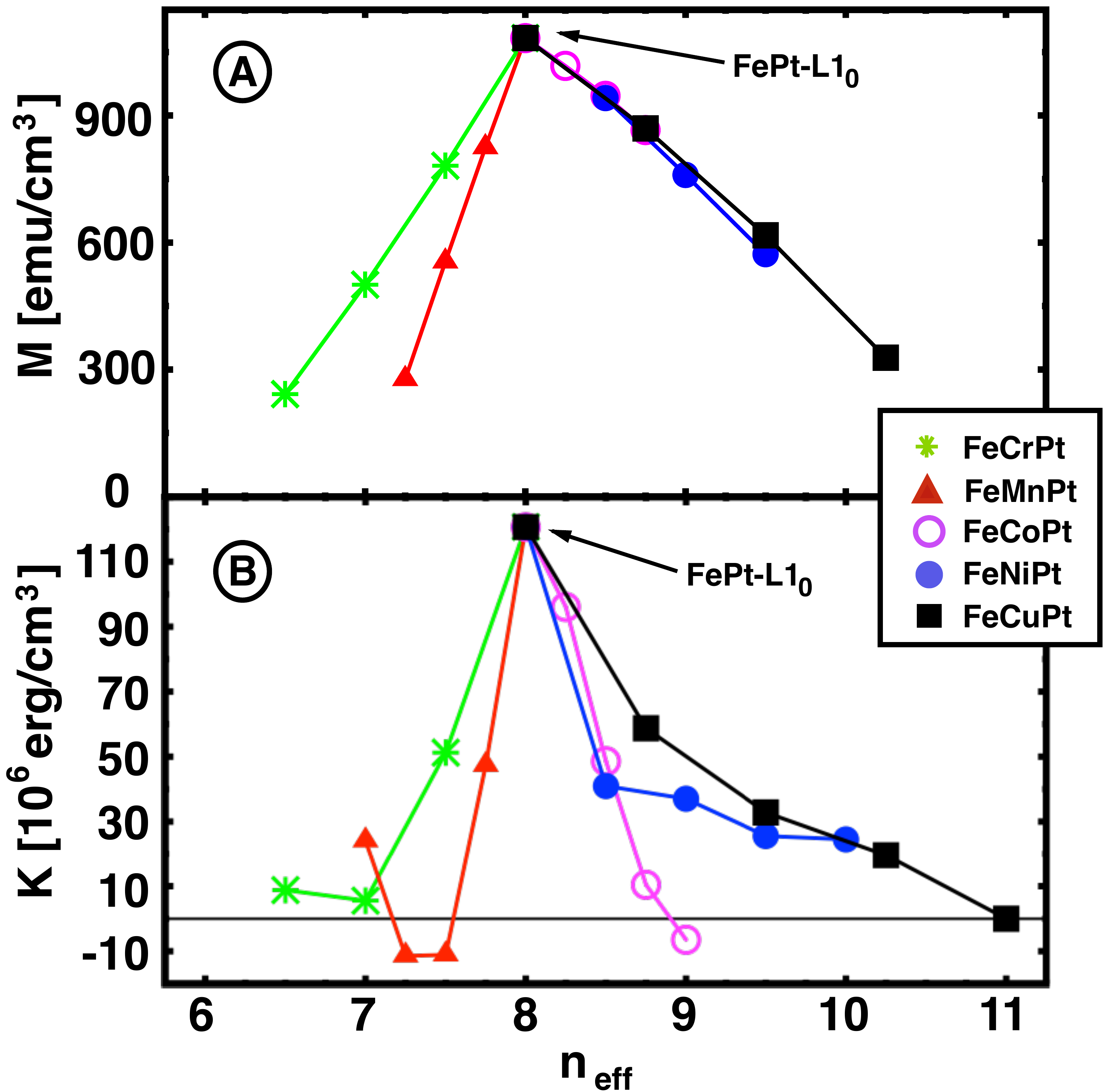}
 \caption{(Color online) Magnetization and magnetic anisotropy energy of 
          Fe$_{1-y}$X$_y$Pt alloys as a function of the $n_{eff}$, A and B, 
          respectively. Peaks on both graphs depict the magnetization and 
          MAE for pure FePt--L1$_0$ alloy.} 
 \label{MAE-fig}
\end{figure}

The effect on magnetization, M, and magnetic anisotropy of X = Cr, Mn, Co, 
Ni, Cu substitution in FePt--L1$_0$ as a function of the $n_{eff}$ is shown 
in Fig.~\ref{MAE-fig}. The variations with $n_{eff}$ of both the magnetization 
and MAE are in good agreement with the experimental data reported by 
Gilbert~\cite{gilbert} et al although a detailed comparison is difficult as 
will be discussed shortly. Consider first the magnetization $\mathbf M$, which 
is calculated taking account of the variation of the MMs between substitution 
sites. It can be seen that  $\mathbf M$ has a maximum for FePt and falls off 
more rapidly for $n_{eff}<8$ than for $n_{eff}>8$. This is consistent with 
experimental data. We note also that the rate of reduction for $n_{eff}<8$ is 
faster than predicted by Sakuma~\cite{sakuma} emphasizing the importance of 
taking specific account of the AF coupling of the impurity spins. The MAE also 
has a maximum for FePt as expected and falls off rapidly on either side of the 
optimum band filling. Consistent with experiment, the dependence of the MAE on 
$n_{eff}$ is slower for Ni and Cu impurities, allowing a more controlled tuning 
of the anisotropy.  Some configurations change their easy axis from out--of--plane 
to  in--plane, specifically in Fe$_{0.50}$Mn$_{0.50}$Pt, Fe$_{0.25}$Mn$_{0.75}$Pt 
and CoPt. This is not observed experimentally, where the range of $n_{eff}$ does 
not extend into the region of the predicted in--plane anisotropy. Finally, we note 
that there is an important dependence of the MAE on the species of the impurity 
atoms, which is not predicted by the (fixed band) calculations of ref~\cite{sakuma}. 
The experimental data cannot reliably be used to test this prediction since the 
results summarised in ref~\cite{gilbert} were all made on different samples using 
different measurement techniques. For example, the MAE for FePtMn differ as much 
as a factor of 2--3 between laboratories, suggesting that the current FePtNi and 
FePtCu data (again measured in different laboratories) cannot be used to test the 
species-dependence predicted here. 

In summary, we have developed a theoretical method to investigate the MAE of the 
FePt--L1$_0$ phase following gradual substitution of Fe by Cr, Mn, Co, Ni and Cu 
keeping the Pt content fixed. The inclusion of the doping elements changes the 
in--plane and the out--of--plane lattice constants characterising the {\it fct} 
phase. In general, {\it a} increases with the reduction of the Fe content promoting 
a decrease of {\it c}. The magnetic moment of the magnetic and non--magnetic species 
also changes substitution. Due to the low Fe--Fe in--plane coordination that emerges 
after replacement of the Fe atoms, the indirect polarization of Pt and other species 
is reduced substantially, disappearing for large dopant concentrations. On the other 
hand, the Fe tends to be magnetically isolated in a non--magnetic environment and 
hence its MM tends to increases. The predicted variation of the magnetization as well 
as the MAE with the effective valence charge  is in good general agreement with prior 
experiments. The calculations also predict that the local, site resolved, anisotropy 
constant has a dispersion arising from differences in the local environment of doping 
atoms situated at different lattice sites. We also predict a species-dependence of the 
variation of MAE with band filling, which requires further experimentation to evaluate, 
but which certainly suggests that fixed band models are insufficient to study the MAE 
of FePtX alloys and that a full treatment of the nature of the alloys is necessary. 

The authors are grateful to Prof. Chih-Huang Lai and Prof Kai Liu for helpful 
discussions. Financial support of the EU Seventh Framework Programme under Grant
No. 281043, FEMTOSPIN is gratefully acknowledged.

\end{document}